\documentclass[prd,preprint,superscriptaddress,amsmath,amssymb,nofootinbib]{revtex4}
\usepackage{graphicx}
\usepackage{dcolumn}
\usepackage{bm}
\usepackage{amssymb}
\usepackage{amsmath}
\usepackage{epsfig}    
\usepackage{color}
\usepackage{slashed}
\usepackage{hhline}

\def\be{\begin{equation}}
\def\ee{\end{equation}}
\newcommand{\bea}{\begin{eqnarray}}
\newcommand{\eea}{\end{eqnarray}}
\newcommand{\nn}{\nonumber}

\begin{document}

\begin{flushright}{KIAS-P17050} \end{flushright}

\title{ Minimal realization of right-handed  gauge symmetry}

\author{Takaaki Nomura}
\email{nomura@kias.re.kr}
\affiliation{School of Physics, KIAS, Seoul 02455, Korea}

\author{Hiroshi Okada}
\email{macokada3hiroshi@cts.nthu.edu.tw}
\affiliation{Physics Division, National Center for Theoretical Sciences, Hsinchu, Taiwan 300}

\date{\today}

\begin{abstract}
 We propose a minimally extended gauge symmetry model with $U(1)_R$, where only the right-handed fermions have nonzero charges in the fermion sector. To achieve both anomaly cancellations and minimality, three right-handed neutrinos are naturally required, and the standard model Higgs has to have nonzero charge under this symmetry.  Then we find that its breaking scale($\Lambda$) is restricted by precise measurement of neutral gauge boson in the standard model; therefore, ${\cal O}$(10) TeV$\lesssim \Lambda$. We also discuss its testability of the new gauge boson and discrimination of $U(1)_R$ model from $U(1)_{B-L}$ one at collider physics such as LHC and ILC.
\end{abstract}
\maketitle

\section{Introduction}
A $U(1)_{B-L}$ gauge symmetry~\cite{Mohapatra:1980qe} is one of the natural extensions of the standard model(SM) to accommodate the three right-handed neutrinos to cancel the gauge anomalies out, and these neutral fermions play roles in arising various sources such as Baryon CP asymmetry of the Universe, a dark matter candidate, as well as light active neutrinos and their mixings with Majorana type, depending on model buildings. Thus a lot of applications have been achieved.

As a similar symmetry that requires three right-handed neutrinos due to the anomaly cancellations~\cite{Ko:2013zsa, Nomura:2016emz},
there exists an right-handed gauge symmetry $U(1)_R$~\cite{Jung:2009jz}, where left-handed fermions do not have nonzero charges under this symmetry.
{The chiral charge assignment for the SM fermions would be natural as the fermions are chiral under $U(1)_Y$ charge assignment. In addition the chiral structure provides richer phenomenology such as forward backward asymmetry at collider experiments.}
However these kinds of models are not applied to a lot of phenomenologies compared to the $B-L$ one,
due to  different charges between right-handed and left-handed fermions.
In particular, a minimal realization of the model with $U(1)_R$ is not sufficiently investigated although the symmetry has been applied to extensions of SM such as two Higgs doublet models~\cite{Ko:2013zsa, Nomura:2016emz,Nomura:2017ezy, Nomura:2016pgg, Ekstedt:2016wyi,Chao:2012mx,Chao:2017rwv}.

In this letter, we realize  a minimal extension of the SM with $U(1)_R$ gauge symmetry, by imposing nonzero $U(1)_R$ charge on the SM Higgs.
{However, in this case, several constraints has to be considered such as oblique parameters~\cite{Peskin:1991sw} and precise measurement of the neutral vector boson of SM ($Z$)~\cite{Olive:2016xmw}.
The most stringent constraint arises from the measurement of the SM $Z$ boson mass ($m_Z$), where the lower bound of vacuum expectation value(VEV) to break the $U(1)_R$ symmetry is ${\cal O}(10)$ TeV~\cite{Nomura:2017tzj}~\footnote{This model naturally leads us to a type-II two Higgs double model. It implies that the constraint from measurement of the SM $Z$ boson mass is more or less the same as our present model due to large hierarchy between two kinds of VEVs.}. 
Therefore} it could still be testable scale at current/future collider experiments.

 This letter is organized as follows.
In Sec. II, {we introduce our model, and formulate Higgs sector, neutral gauge sector, neutrino sector, and interacting terms.
And we {discuss} phenomenologies of new neutral gauge boson at colliders.
 Finally we devote the summary of our results and the conclusion.}

\section{Model setup and Constraints}
\begin{table}[t!]
\begin{tabular}{|c||c|c|c|c|c|c|c|c|}\hline\hline  
& ~$Q_L^a$~& ~$u_R^a$~  & ~$d_R^a$~& ~$L_L^a$~& ~$e_R^a$~& ~$\nu_R^a$~& ~$H$~& ~$\varphi$~ \\\hline\hline 
$SU(3)_C$ & $\bm{3}$  & $\bm{3}$ & $\bm{3}$ & $\bm{1}$ & $\bm{1}$ & $\bm{1}$ & $\bm{1}$ & $\bm{1}$  \\\hline 
$SU(2)_L$ & $\bm{2}$  & $\bm{1}$  & $\bm{1}$  & $\bm{2}$  & $\bm{1}$  & $\bm{1}$  & $\bm{2}$ & $\bm{1}$   \\\hline 
$U(1)_Y$   & $\frac16$ & $\frac23$ & $-\frac13$ & $-\frac12$  & $-1$ & $0$  & $\frac12$  & $0$\\\hline
$U(1)_{R}$   & $0$ & $x$ & $-x$   & $0$  & $-x$  & $x$  & $x$  & $2x$\\\hline
\end{tabular}
\caption{ 
Charge assignments of the our fields
under $SU(3)_C\times SU(2)_L\times U(1)_Y\times U(1)_{R}$ with $x\neq0$, where its upper index $a$ is the number of family that runs over $1-3$.}
\label{tab:1}
\end{table}

In this section we formulate our model and derive some formulas such as $Z$-$Z'$ mixing and neutrino mass matrix.
We add three families of right-handed Majorana fermions $\nu_R^a$($a=1-3$) and an isospin singlet boson $\varphi$, both of which carry  
nonzero $U(1)_R$ charges. Furthermore, the SM Higgs boson also has nonzero $U(1)_R$ charge that plays an crucial role in determining the scale of $U(1)_R$ breaking 
$\Lambda \gtrsim {\cal O}$(10) TeV as discussed later. All the field contents and their assignments are summarized in Table~\ref{tab:1}.
The relevant Yukawa interactions and scalar potential under these symmetries is given by 
\begin{align}
&-{\cal L}_{Y}
=  (y_u)_{ab} \bar Q^a_L \tilde H u^b_R +  (y_d)_{ab} \bar Q^a_L H d^b_R+  (y_\ell)_{ab} \bar L^a_L H e^b_R\nn\\
& \qquad \quad +  (y_D)_{ab} \bar L^a_L\tilde H \nu^b_R  +  (y_\nu)_{aa} \bar\nu^{aC}_R \nu^a_R \varphi^*
+ {\rm h.c.}, \label{Eq:yuk} \\
& {\cal V}
= -\mu^2_1 |\varphi|^2 -\mu^2_2 |H|^2 
+ \lambda_1 |\varphi|^4 + \lambda_2 |H|^4  + \lambda_{3} |\varphi|^2 |H|^2 ,
\label{Eq:pot}
\end{align}
where $\tilde H\equiv i\sigma_2H$, and upper indices $(a,b)=1$-$3$ are the number of families, and $y_\nu$ can be diagonal matrix without loss of generality
due to the phase redefinitions of fermion fields.
The masses of the SM fermions in both quark and charged-lepton sector are given by $m_u=y_u v/\sqrt2$, $m_d=y_d v/\sqrt2$, and $m_\ell=y_\ell v/\sqrt2$, which are the same as the SM one.   
{Notice that our Lagrangian has accidental global symmetry of lepton number when we assign lepton number 2 for $\varphi$ which will be broken after $\varphi$ developing a VEV. Also there is an accidental global baryon number symmetry as in the SM.}

{\it Scalar sector}:
The scalar fields are parameterized as 
\begin{align}
&H =\left[\begin{array}{c}
w^+\\
\frac{v + r +i z}{\sqrt2}
\end{array}\right],\quad 
\varphi=
\frac{v'+ r' + iz' }{\sqrt2},
\label{component}
\end{align}
where $w^+$, $z$, and $z'$ are massless Nambu-Goldstone(NG) bosons which are absorbed by the SM gauge bosons $W^+$ and $Z$, and extra $Z'$ boson from $U(1)_R$.  
Inserting tadpole conditions for $r$ and $r'$, we obtain the mass matrix for CP even scalar, $m_R^2$, in the basis of $(r,r')$, and 
the mass eigenstates $\{ h,H \}$ is found to be $(r,r')^T=O_R^T (h, H)^T$,
where mass eigenvalues are given by $m_{h,H}^2=O_R m_R^2 O_R^T$; $m_R^2$ and $O_R$ are obtained as
\begin{align}
m_{R}^2
&=
\left[\begin{array}{cc}
2 v^2\lambda_{2} &  v v' \lambda_{3}  \\ 
v v' \lambda_{3} & 2 v'^2 \lambda_{1} \\ 
\end{array}\right],
O_R
=
\left[\begin{array}{cc}
-c_\theta &  s_\theta  \\ 
s_\theta &  c_\theta \\ 
\end{array}\right],
\end{align}
with $s_{2\theta}=\frac{2vv'\lambda_3}{m^2_{h}-m^2_{H}}$. 
The mass eigenvalues are also calculated such that
\begin{equation}
m_{h,H}^2 = (v^2 \lambda_2 + v'^2 \lambda_1) \mp \sqrt{(v^2 \lambda_2 - v'^2 \lambda_1)^2 + v^2 v'^2 \lambda_3^2}.
\end{equation}
Here $h_1\equiv h_{SM}$ is the SM Higgs, therefore,  $m_{h}=$125 GeV.
The mixing effect for CP-even scalar is constrained by the measurements of Higgs production cross section and its decay branching ratio at the LHC, 
and $s_a\lesssim 0.2$ is provided by the current data~\cite{Olive:2016xmw}.

{\it $Z_{SM}-Z'$ mixing}:
Since $H$ has nonzero $U(1)_R$ charge, there is mixing between $Z_{SM}$ and $Z'$. 
Taking $x=1$~\footnote{For $x \neq 1$,  we can obtain constraints on the gauge coupling by scaling as $g'/x$ from our analysis with $x=1$.}, the resulting mass matrix in basis of $(Z_{SM},Z')$  is given by
\begin{align}
m_{Z_{SM}Z'}^2
&= \frac14
\left[\begin{array}{cc}
(g_1^2+g_2^2) v^2 &  -2\sqrt{g_1^2+g_2^2}g' v^2  \\ 
-2 \sqrt{g_1^2+g_2^2}g' v^2  & 4 g'^2 (v^2+ 4 v'^2)   \\ 
\end{array}\right]
=
m_{Z'}^2
\left[\begin{array}{cc}
\epsilon_1^2 & -\epsilon_1 \epsilon_2  \\ 
-\epsilon_1 \epsilon_2 & 1+\epsilon_2^2  \\ 
\end{array}\right],
\end{align}  
where $m_{Z_{SM}}\equiv \frac{\sqrt{g_1^2+g_2^2}v}{2}\approx 91.18$ GeV, $m_{Z'}\equiv 2 g'v'$, $\epsilon_1\equiv \frac{m_{Z_{SM}}}{m_{Z'}}$, $\epsilon_2\equiv \frac{v}{2v'}$, $g_1$, $g_2$, and $g'$ are gauge coupling of $U(1)_Y$, $SU(2)_L$, and $U(1)_R$, respectively.
Then its mass matrix is diagonalized by the two by two mixing matrix $V$ as $V m_{Z_{SM}Z'}^2 V^T
\equiv {\rm Diag}(m^2_{Z_{}},m^2_{Z_{R}}) $,
where we work under $\epsilon_2^2<<1$ and
\begin{align}
m^2_{Z}&\approx m_{Z_{SM}}^2(1-\epsilon_2^2),\
m^2_{Z_{R}}\approx m_{Z'}^2 (1 +  \epsilon_1^2\epsilon_2^2),\label{eq:zm}
\\
V&\approx
\left[\begin{array}{cc}
c_{Z} &  s_{Z} \\ 
-s_{Z}  &  c_{Z}  \\ 
\end{array}\right], \quad \theta_{Z} = \frac{1}{2} \tan^{-1} \left[ \frac{2 \epsilon_1 \epsilon_2}{1+\epsilon_2^2-\epsilon_1^2} \right].
\end{align} 
Since the ambiguity of the $Z$ boson mass is around $0.0021$~\cite{Olive:2016xmw} we require
\begin{align}
|\Delta m_Z|=m_{Z_{SM}}\left(\sqrt{1-\epsilon_2^2}-1\right)\lesssim 0.0021\ {\rm GeV}.
\label{eq:const_zm}
\end{align}       
Therefore one finds the stringent constraint on the $v'$ from Eq.(\ref{eq:zm}) and (\ref{eq:const_zm}) as
\begin{align}
18.13\ {\rm TeV}\lesssim v'.
\label{eq:limit_v}
\end{align}    
{The $Z_R$ mass can be approximated as $m_{Z_R} \simeq m_{Z'} = 2 g' v'$ since $\epsilon_{1,2}$ is small. Thus gauge coupling $g'$ is almost fixed if we choose values of $m_{Z_R}$ and $v'$. }

{\it Fermion sector}:
Here we focus on the neutral sector, since the other three sectors are the same the SM one as discussed above.
The six by six mass mass matrix in basis of $(\nu_L,\nu_R)$ is given by
\begin{align}
{\cal M}_N=
\left[\begin{array}{cc}
0 & m_D  \\ 
m_D^T  &  M  \\ 
\end{array}\right],
\end{align}
and ${\cal M}_N$ is diagonalized by $(D_{\nu_l},D_{\nu_H})\equiv O_N {\cal M}_N O_N^T$,
where $m_D\equiv y_D v/\sqrt2$,  $M\equiv y_\nu v'/\sqrt2$, and $O_N$ is six by six unitary matrix. 
Assuming $m_D<<M$, one finds the following mass eigenvalues and their mixing $O_N$:
\begin{align}
 D_{\nu_l} & \equiv V_{MNS} m_\nu V^T_{MNS}
 \approx -2V_{MNS} m_D M^{-1} m_D^T V^T_{MNS} , \\
D_{\nu_H} & \approx M,\quad 
O_N\approx
\left[\begin{array}{cc}
V_{MNS} & 0  \\ 
0  &  1  \\ 
\end{array}\right]
\left[\begin{array}{cc}
-1 & \theta  \\ 
\theta^T  &  1  \\ 
\end{array}\right],
\end{align}
where $\theta\equiv m_D M^{-1}$, $V_{MNS}$ and $D_{\nu_l}$ are observable and fixed by the current neutrino oscillation data~\cite{Olive:2016xmw}.
Supposing $y_\ell={\cal O}$(1) and $v\approx 246$ GeV and $v'\approx 18$ TeV,
one finds the typical order of $y_D$ is $10^{-5}$ in order to reproduce the observed  neutrino mass scale of $\lesssim 0.1$ eV.
The scale of $y_D$ can be tested by the current experimental data at the LHC~\cite{Huitu:2008gf}.
One also finds the following relation between flavor- and mass-eigenstate:
\begin{align}
\nu_L\approx -V^T_{MNS} \nu_l + \theta \nu_H,\quad
\nu_R\approx -\theta^\dag V^\dag_{MNS} \nu_l + \nu_H.
\end{align}

{\it Interactions via kinetic terms}:
Now that we formulate the masses and their mixings for the neutral fermions,
one can write down the interacting term from the kinetic Lagrangian under $SU(2)_L\times U(1)_Y\times U(1)_R$ as:
\begin{align} 
{\cal L}_{}&\sim
\frac{g_2}{\sqrt2}
\sum_i^3\left[W^-_\mu\bar\ell_a \gamma^\mu P_L(-(V_{MNS}^T)_{ai} \nu_{l_i} +\theta_{ai} \nu_{H_i} ) +{\rm h.c.}\right]\nn\\
&-\frac{g_2}{2 c_w}\sum_{i,j=1}^3 Z_\mu
\left[ (V^*_{MNS})_{ja} \theta_{ai} \bar\nu_{l_j} \gamma^\mu P_R \nu_{H_i} + {\rm h.c.}\right]\nn\\
&-xg' \sum_{j=1}^3 Z'_\mu
\left[ (V_{MNS}\theta)_{ja} \bar\nu_{l_j} \gamma^\mu P_R \nu_{H_j} + {\rm h.c.}\right],
\end{align}
where $c_w\equiv \cos \theta_w$ is Weinberg angle.
The other three sectors are given by
\begin{align} 
{\cal L}_{}\sim
&-xg' (Z'_\mu - \epsilon_2^2 Z_{\mu} ) 
 \times \left[ \bar u_a\gamma^\mu P_R u_a -  \bar d_a\gamma^\mu P_R d_a - \bar \ell_a\gamma^\mu P_R \ell_a \right].
\end{align}
In the following, we take $x=1$ for simplicity.

\section{$Z'$ physics at colliders}
Here we discuss collider physics of $Z'$ boson. Since $Z'$ couples to the SM leptons, the LEP experiment provides the constraint on the effective interaction induced from $Z'$ exchange;
\begin{equation}
{\cal L}_{eff} = \frac{1}{1+ \delta_{e \ell}} \frac{g'^2}{m_{Z'}^2} (\bar e \gamma^\mu P_R e)(\bar f \gamma_\mu P_R f), \label{eq:eff}
\end{equation}
where $f$ indicates all the fermions in the model.
Note that the above interaction includes only the right-handed chirality
 due to the nature of $U(1)_R$ symmetry.
The analysis of data by LEP experiment in Ref.~\cite{Schael:2013ita} provides the constraint such that
\begin{equation}
\frac{m_{Z'}}{g'} \gtrsim 3.7 \ {\rm TeV}.
\end{equation}
We thus find that constraint from $Z_{SM}$-$Z'$ mixing gives more stringent bound than the one from LEP
for $m_{Z'}/g' = 2 v'$ as in Eq.~(\ref{eq:limit_v}).

\begin{figure}[t]
\begin{center}
\includegraphics[width=80mm]{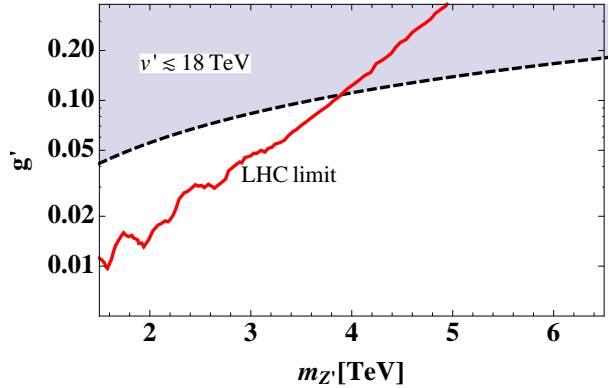} \qquad
\caption{The current LHC limit on $\{m_{Z'}, g'\}$ plane is shown where region above red curve is excluded by the latest data~\cite{ATLAS:2017wce}. The gray region is constrained by Eq.~(\ref{eq:limit_v}) from the relation $g'/m_{Z'} = 1/(2v')$.} 
  \label{fig:ZpLHC}
\end{center}\end{figure}
\begin{figure}[t]
\begin{center}
\includegraphics[width=80mm]{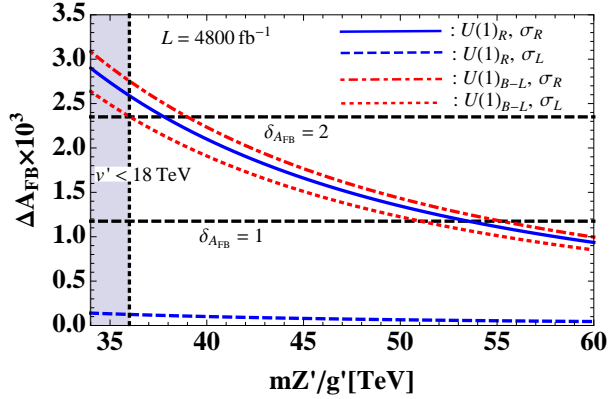} \qquad
\caption{$\Delta A_{FB}$ defined as Eq.~(\ref{eq:delAFB}) as a function of $m_{Z'}/g'$ where we apply polarized cross section for the $Z'$ from $U(1)_R$ and $U(1)_{B-L}$. Here the horizontal lines indicate $\delta_{A_{FB}} = 1$ and 2 corresponding to sensitivity confidence level $1 \sigma$ and $2 \sigma$, and the vertical line corresponds to $v' = 18$ TeV.} 
  \label{fig:AFB}
\end{center}\end{figure}

{The $Z'$ boson can be produced at the LHC via the process $q \bar q \to Z'$ and decay channel of $Z' \to \ell^+ \ell^- (\ell = e, \mu)$ will provide the most significant signature.
The branching fraction for $Z' \to \ell^+ \ell^-$ is similar to $U(1)_{B-L}$ case since the $Z'$ universally couples to quarks and leptons. We estimate the cross section with {\it CalcHEP}~\cite{Belyaev:2012qa} implementing relevant interactions and using the CTEQ6 parton distribution functions (PDFs)
~\cite{Nadolsky:2008zw}. Then we obtain the value of $6 \times 10^{-5}$ pb for the process $pp \to Z' \to \mu^+ \mu^-$ with $m_{Z'} =$ 4 TeV and $g' =0.1$. The Fig.~\ref{fig:ZpLHC} shows the excluded region by the current LHC data~\cite{ATLAS:2017wce} and the VEV constraint in Eq.~(\ref{eq:limit_v}). We find that the current LHC constraint is stronger(weaker) than that from $v'$ for $m_{Z'} \lesssim(\gtrsim) \ 3.9$ TeV, and wider parameter region of $\{g', m_{Z'} \}$ can be tested by the future LHC experiments. The chirality structure could be tested at the LHC by measuring  forward-backward and top polarization asymmetries in $Z' \to t \bar t$ mode~\cite{Cerrito:2016qig}, which will distinguish our $Z'$ interaction from the other $Z'$ interactions like $U(1)_{B-L}$.

The effective interaction in Eq.~(\ref{eq:eff}) can also be tested by measuring the process $e^+ e^- \to f \bar f$ at the International Linear Collider (ILC) even if $Z'$ mass is very heavy to directly produce at the LHC. 
In particular, analysis with polarized initial state is useful to distinguish our model from other $Z'$ such as $U(1)_{B-L}$, since our $Z'$ couples only right-handed SM fermions. The partially-polarized differential cross section can be written by~\cite{Nomura:2017abh}
\begin{align}
\frac{d \sigma (P_{e^-}, P_{e^+})}{d \cos \theta} = \sum_{\sigma_{e^-}, \sigma_{e^+} = \pm} \frac{1+ \sigma_{e^-} P_{e^-}}{2} \frac{1 +\sigma_{e^+} P_{e^-}}{2} \frac{d \sigma_{\sigma_{e^-} \sigma_{e^+}}}{d \cos \theta},
\end{align}
where $P_{e^-(e^+)}$ is the degree of polarization for the electron(positron) beam and $\sigma_{\sigma_{e^-} \sigma_{e^+}}$ indicates the cross section when the helicity of initial electron(positron) is $\sigma_{e^- (e^+)}$; the helicity of final states is summed up and more detailed form is found in ref~\cite{Nomura:2017abh}. We then define polarized cross sections $\sigma_{L,R}$ by following two cases as realistic values at the ILC~\cite{Baer:2013cma}:
\begin{equation}
\frac{d \sigma_{R}}{d \cos \theta} = \frac{d \sigma (0.8,-0.3)}{d \cos \theta}, \quad \frac{d \sigma_{L}}{d \cos \theta} = \frac{d \sigma (-0.8,0.3)}{d \cos \theta}.
\end{equation}
The polarized cross sections are applied to study the sensitivity to $Z'$ through the measurement of a forward-backward asymmetry at the ILC which is given by 
\begin{align}
& A_{FB} = \frac{N_F - N_B}{N_F + N_B}, \nonumber \\
& N_{F(B)} = \epsilon L \int_{0(-c_{\rm max})}^{c_{\rm max}(0)} d \cos \theta \frac{d \sigma}{d \cos \theta},
\end{align}
where a kinematical cut $c_{max}$ is chosen to maximize the sensitivity, $L$ is an integrated luminosity and $\epsilon$ is an efficiency depending on the final states.
Here we assume $\epsilon = 1$ for electron and muon final states, and $c_{\rm max} = 0.5(0.95)$ is taken for electron(muon) final state~\cite{Tran:2015nxa}.
Then we estimate the forward-backward asymmetry for cases with only SM gauge boson contributions, and with both SM and $Z'$ boson contributions, in order to investigate the sensitivity to $Z'$; the former case gives $N_{F(B)}^{SM}$ and $A_{FB}^{SM}$ while the latter case $N_{F(B)}^{SM+Z'}$ and $A_{FB}^{SM+Z'}$. Furthermore, we consider two types of $Z'$ boson interaction with SM-fermions; vector-like couplings and right-handed chiral couplings which are respectively correspond to $U(1)_{B-L}$ and $U(1)_{R}$ cases.   
The sensitivity to $Z'$ contribution is estimated by 
\begin{equation}
\Delta A_{FB} = |A_{FB}^{SM+Z'}- A_{FB}^{SM}|,
\label{eq:delAFB}
\end{equation}
which is compared with a statistical error of the asymmetry, assuming only SM contribution
\begin{equation}
\delta_{A_{FB}} = \sqrt{\frac{1-(A_{FB}^{SM})^2}{N_F^{SM}+N_B^{SM}}}.
\end{equation}
Here we focus on the $\mu^+ \mu^-$ mode since it is the most sensitive one~\cite{Nomura:2017tzj}.
In Fig.~\ref{fig:AFB}, we show $\Delta A_{FB}$ for polarized cross sections of the $e^+ e^- \to \mu^+ \mu^-$ process as a function of $m_{Z'}/g'$, where we have considered $Z'$ from both $U(1)_{R}$ and $U(1)_{B-L}$ cases for comparison.
The integrated luminosity of $L = 4800$ fb$^{-1}$ is applied in the figure which is expected in the upgraded ILC~\cite{Barklow:2015tja}, and horizontal lines indicate $\delta_{A_{FB}} = 1$ and $2$ corresponding to $1 \sigma$ and $2 \sigma$ sensitivity confidence level.
We find that region with $m_{Z'}/g' \lesssim 38$ TeV can give more than $2 \sigma$ sensitivity by measuring only $\mu^+ \mu^-$ mode.
{\it Remarkably, the $U(1)_R$ case provides significant difference of $\Delta A_{FB}$ for $\sigma_R$ and $\sigma_L$  due to the chirality structure of $Z'$ interaction from $U(1)_R$, while $U(1)_{B-L}$ case gives small difference.}
Thus we can distinguish the $Z'$ interactions from $U(1)_{R}$ and $U(1)_{B-L}$ by comparing the $\Delta A_{FB}$ from polarized cross sections $\sigma_{L}$ and $\sigma_R$.
The higher sensitivity can also be expected to combining different final states such as other leptonic modes $\{ e^+ e^-, \tau^+ \tau^- \}$ as well as hadronic modes $jj$.
Moreover, detailed analysis of fitting the scattering angular distribution will help us to enhance the sensitivity to $Z'$ interaction~\cite{ilc}, although it is beyond the scope of our paper.
In addition, more precise measurement of $Z_{SM}$ mass at the ILC will also test our model.}

{More phenomenology can be considered when the heavy neutrinos are not very heavy since $pp \to Z' \to \nu_H \nu_H$ process would be realized at the LHC. 
Heavy neutrinos in the final state will decay into $\ell^\pm W^\mp$ through mixing effect $\theta$, and the invariant mass of final state provides $Z'$ mass. A detailed analysis of the process will be given elsewhere.   }

\section{Summary and Conclusions}
 We have proposed a minimally extended $U(1)$ gauge symmetry, where only right-handed fermions have nonzero charges in the fermion sector.
 To achieve anomaly cancellations, three right-handed neutrinos are naturally requested where anomaly cancels among each of family.
To realize our model minimally, we have imposed nonzero charge under this symmetry for the standard model Higgs.
 Then we have found that its breaking scale($\Lambda\sim v'$) is restricted by precise measurement of the standard model neutral gauge boson; therefore, 18 TeV$\lesssim v'$.
  
We have discussed implications of our $Z'$ boson to collider physics considering the LEP constraint, $Z'$ production at the LHC and $e^+ e^- \to f \bar f$ process at the ILC.
The LEP constraint is weaker than the constraint from $Z_{SM}$-$Z'$ mixing while the LHC can explore wider parameter region of $\{g', m_{Z'} \}$ in current/future experiments.
The parameter region with $v'$ beyond the $Z_{SM}$-$Z'$ mixing constraint can be tested by the ILC experiment with sufficient integrated luminosity. Moreover we can test the chirality structure of the interaction with polarized electron(positron) beam, and our $Z'$ could be discriminated from the other $Z'$ such as $U(1)_{B-L}$ gauge symmetry. \\


\section*{Acknowledgments}
H. O. is sincerely grateful for KIAS and all the members.

\end{document}